\documentclass[11pt]{article}
\usepackage{latexsym}
\usepackage{amsmath}
\usepackage{amssymb}
\usepackage{amsfonts}
\usepackage{amsthm}
\usepackage{url}
\usepackage{graphicx}
\usepackage{isolatin1}
\usepackage{subfigure}
\usepackage{rotating}
\usepackage{epsfig}
\usepackage{times} 
\usepackage[Algorithm]{algorithm}
\usepackage[noend]{algorithmic}
\usepackage{setspace}

\newtheorem{theorem}{Theorem}[section]
\newtheorem{lemma}{Lemma}[section]

\def\R{{\mathbb R}}

\def\qed{\hfill$\diamond$}
\def\Prob{{\mathbb P}}
\def\Sim{\mbox{\rm Sim}}

\makeatletter
\def\@seccntformat#1{}
\makeatother

\begin{document}

\title{Pair HMM based gap statistics for re-evaluation of indels in alignments with affine gap penalties: Extended Version}

\author{Alexander Sch{\"o}nhuth$^1$, Raheleh
  Salari$^2$, S.~Cenk Sahinalp$^2$\\[2ex]
$^1$ Department of Mathematics, University of California, Berkeley\\
$^2$ School of Computing Science, Simon Fraser University, Burnaby
}
\maketitle

\begin{abstract}
Although computationally aligning sequence is a crucial step in the
vast majority of comparative genomics studies our understanding of
alignment biases still needs to be improved. To infer true structural
or homologous regions computational alignments need further
evaluation. It has been shown that the accuracy of aligned positions
can drop substantially in particular around gaps. Here we focus on
re-evaluation of score-based alignments with affine gap penalty costs.
We exploit their relationships with pair hidden Markov models and
develop efficient algorithms by which to identify gaps which are
significant in terms of length and multiplicity. We evaluate our
statistics with respect to the well-established structural alignments
from SABmark and find that indel reliability substantially increases
with their significance in particular in worst-case twilight zone
alignments. This points out that our statistics can reliably
complement other methods which mostly focus on the reliability of
match positions.
\end{abstract}

\section{Introduction}

Having been introduced over three decades ago \cite{Needleman70}
the sequence-alignment problem has remained one of the most actively
studied topics in computational biology. While the vast majority of
comparative genomics studies crucially depend on alignment quality
inaccuracies abundantly occur. This can have detrimental effects in
all kinds of downstream analyses \cite{Loeytynoja08}. Still, our
understanding of the involved biases remains rather rudimentary
\cite{Kumar07},\cite{Lunter07}. That different methods often yield
contradictory statements \cite{Dewey06} further establishes the need
for further investigations into the essence of alignment biases and
their consequences \cite{Kumar07}.\par While the sequence-alignment
problem virtually is that of inferring the correct placement of gaps,
insertions and deletions (indels) have remained the most unreliable
parts of the alignments. For example, Lunter et al. \cite{Lunter07},
in a whole-genome alignment study, observe 96\% alignment accuracy for
alignment positions which are far away from gaps while accuracy drops
down to 56\% when considering positions closely surrounding gaps.
They also observe a downward bias in the number of inferred indels
which is due to effects termed gap attraction and gap
annihilation. Decreased numbers of inferred indels were equally
observed in other recent studies
\cite{Loeytynoja05,Polyanovsky08}. This points out that {\em numbers
  and size} of computationally inferred indels can make statements
about alignment quality.\par The {\em purpose of this paper} is to
{\em systematically address} such questions. We develop a statistical
framework by which to efficiently compute probabilities of the type
\begin{equation}
\label{eq.ilp}
\Prob(I_{d,\mathcal{A}}(x,y)\ge k\,|\,L_{\mathcal{A}}(x,y)=n,\Sim_{\mathcal{A}}(x,y)\in[\sigma_1,\sigma_2])
\end{equation}
where $(x,y)$ has been randomly sampled from an appropriate pool of
protein pairs. In the following pools contain protein pairs which have
a (either false or true positive) structural SABMark \cite{VanWalle05}
(see below) alignment. In case of, for example, all pairs of human
proteins, (\ref{eq.ilp}) would act as null distribution for human.
$\mathcal{A}$ is a local or global optimal, score-based alignment
procedure with affine gap penalties such as the affine gap cost
version of the Needleman-Wunsch (NW) algorithm
\cite{Needleman70,Gotoh82} or the Smith-Waterman (SW) algorithm
\cite{Smith81,Waterman87}, $L_{\mathcal{A}}(x,y)$ is the length of the
alignment, $\Sim_{\mathcal{A}}(x,y)$ denotes alignment similarity that
is the fraction of perfectly matching and ``well-behaved'' mismatches
vs.~''bad'' mismatches (as measured in terms of biochemical affinity
\cite{Pearson88}) and gap positions. $I_{d,\mathcal{A}}(x,y)$ finally
denotes the length of the $d$-th longest gap in the alignment. In
summary, (\ref{eq.ilp}) can be read as the probability that a NW
resp.~SW alignment of length $n$ and similarity between $\sigma_1$ and
$\sigma_2$ contains at least $d$ gaps of length $k$ and the reasoning
is that gaps which make part of significant such gap combinations are
more likely to reflect true indels. Significance is determined
conditioned the length $L(x,y)$ of the alignment as well as alignment
similarity $\Sim(x,y)$. The reasoning behind this is that longer
alignments are more likely to accumulate spurious indels such that
only increased gap length and multiplicity are significant signs of
true indels. Increased similarity $\Sim(x,y)$, however, indicates that
already shorter and less gaps are more likely to reflect true indels
simply because an alignment of high similarity is an overall more
trustworthy statement. In summary, we provide a {\em statistically
  sound, systematic} approach to answering questions such as ``Am I to
believe that $4$ gaps of size at least $6$ in an alignment of length
$200$ and similarity $50$ are likely to reflect true indels'' as
motivated by the recent studies
\cite{Loeytynoja05,Lunter07,Polyanovsky08}.\par We opted to address
these questions for score-based alignments with affine gap costs for
two reasons:\par
\begin{enumerate}
\item To employ score-based such alignments still is a most popular option
  among most bioinformatics practitioners.\par
\item Such alignments can be alternatively viewed as Viterbi paths in 
  pair HMMs. While exact statistics on Viterbi paths are hard to obtain
  and beyond the scope of this study we obtain reasonable approximations
  by ``Viterbi training'' sensibly modified versions of the hidden Markov
  chains which underlie the pair HMMs.\par
\end{enumerate}
We evaluate our statistics on the well-established SABmark
\cite{VanWalle05} alignments. SABmark is a database of structurally
related proteins which cover the entire known fold space. The
``Twilight Zone set'' was particularly designed to represent the worst
case scenario for sequence alignment.  While we obtain good results
also in the more benign ``Superfamilies set'' of alignments it is that
worst case scenario of twilight zone alignments where our statistics
prove their particular usefulness.  Here significance of gap
multiplicity is crucial while significance of indel length alone does
not necessarily indicate enhanced indel quality.

\subsection{Related Work}

\cite{Mevissen96} re-evaluate match (but not indel) positions in
global score-based alignments by obtaining reliability scores from
suboptimal alignments. Similarly, \cite{Schlosshauer02} derive
reliability scores also for indel positions in global score-based
alignments.  However, the method presented in \cite{Schlosshauer02}
reportedly only works in the case of more than 30\% sequence identity.
Related work where structural profile information is used is
\cite{Tress03} whereas \cite{Cline02} re-align rather than
re-evaluate.\par Posterior decoding algorithms (see
e.g.~\cite{Do05,Lunter07,Bradley09} for most recent approaches) are
related to re-evaluation of alignments insofar as posterior
probabilities can be interpreted as reliability scores. However, how
to score indels as a whole by way of posterior decoding does not have
a straightforward answer. We are aware of the potential advantages
inherent to posterior decoding algorithms---it is work in progress of
ours to combine the ideas of pair HMM based posterior decoding
aligners with the ideas from this study\footnote{Note that although we
  derive statistical scores for indels as a whole our evaluation in
  the Results section will refer to counting individual indel
  positions.}.\par To assess statistical significance of alignment
phenomena is certainly related to the vastly used
Altschul-Dembo-Karlin statistics \cite{Karlin90,Dembo91,Altschul96}
where score significance serves as an indicator of protein
homology.\par To devise computational indel models still remains an
area of active research
(e.g.~\cite{Qian01,Chang04,Cartwright06,Lunter07,Miklos08}). However,
the community has not yet come to a final conclusion.\par Last but not
least, the algorithms presented here are related to the algorithms
developed in \cite{Schoenhuth10} where the special case of $d=1$ for
only global alignments in (\ref{eq.milp}) was treated to explore the
relationship of indel length and functional divergence. The advances
achieved here are to provide null models also for the more complex
case of local alignments and to devise a dynamic programming approach
also for the case $d>1$ which required to develop generalized
inclusion-exclusion arguments.\par Just like in \cite{Schoenhuth10}
note that {\it empirical statistics approaches fail} for the same
reasons that have justified the development of the
Altschul-Dembo-Karlin statistics: sizes of samples are usually much
too small. Here samples (indels in alignments) are subdivided into
bins of equal alignment similarity and then further into bins of equal
length $n$ and $d$-th longest indel size $k$.

\subsection{Summary of Contributions}

As above-mentioned, our work is
centered around computation of probabilities
\begin{equation}
\label{eq.milp}
\Prob(I_{d}(x,y)\ge k\,|\,L(x,y)=n,\Sim(x,y)\in [\sigma_1,\sigma_2]).
\end{equation}
We refer to this problem as {\em Multiple Indel Length Problem (MILP)}
in the following. Our contributions then are as follows:\par
{\bf 1.} We are the first ones to address this problem and derive
  appropriate Markov chain based null models from the pair HMMs which
  underlie the NW resp.~SW algorithms to yield {\em approximations}
  for the probabilities (\ref{eq.milp}).\par
{\bf 2.} Despite having a natural formulation, the inherent Markov chain
  problem had no known efficient solution. We present the first
  efficient algorithm to solve it.\par
{\bf 3.} We demonstrate the usefulness of such statistics by showing that
  significant gaps in both global and local alignments indicate
  increased reliability in terms of identifying true structural
  indel positions. This became particularly obvious for worst-case twilight
  zone alignments of at most 25\% sequence identity.\par
{\bf 4.} Thereby we deliver statistical evidence of that computational
  alignments are biased in terms of numbers and sizes of gaps as
  described in \cite{Lunter07,Polyanovsky08}. In particular too little
  numbers of gaps can reflect alignment artifacts.\par
{\bf 5.} Re-evaluation of indels in score-based both local and global
  alignments had not been explicitly addressed before, in particular,
  reliable solutions for worst-case twilight zone alignments were
  missing. Our work adds to (rather than competes with) the above-mentioned
  related work.

In {\em summary}, we have complemented extant methods for score-based
alignment re-evaluation. Note that none of the existing methods
explicitly addresses indel reliability but rather focus on the
reliability of substitutions.

\section{Methods}\label{sec.indstat}

\subsection{Pair HMMs and Viterbi Path Statistics}
\label{ssec.approx}

\begin{figure*}[!tbp]
\vspace{-0.1in}
 \begin{center}
 \subfigure[Standard pair HMM]
  {\includegraphics[width=0.8\columnwidth]{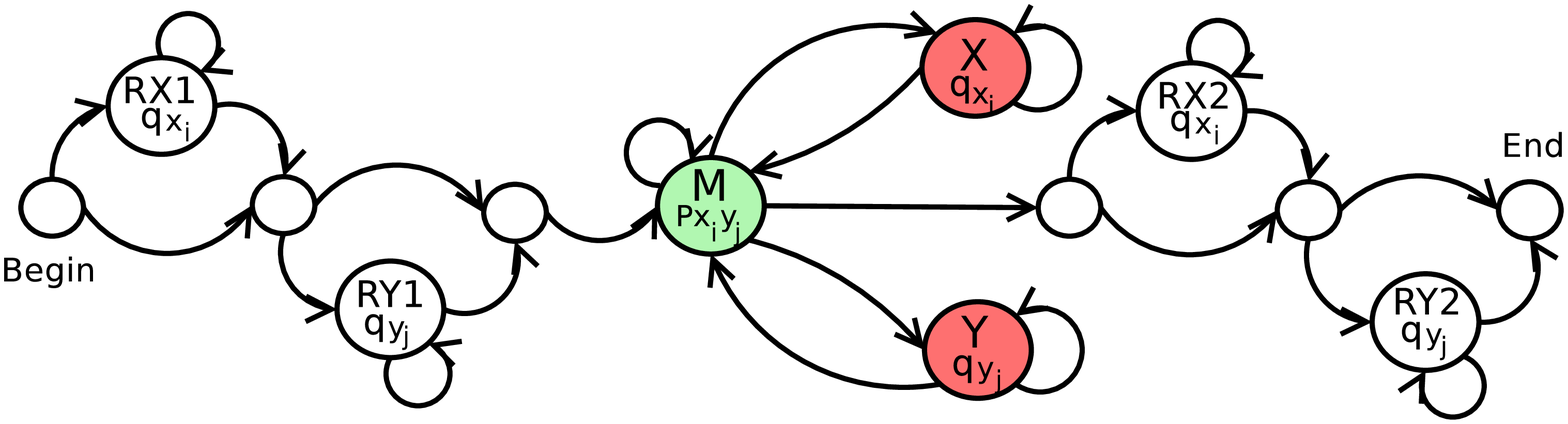}\hspace{.5cm}
   \label{fig.pairhmm}
   }
 \subfigure[Markov Chain]
   {\includegraphics[width=0.65\columnwidth]{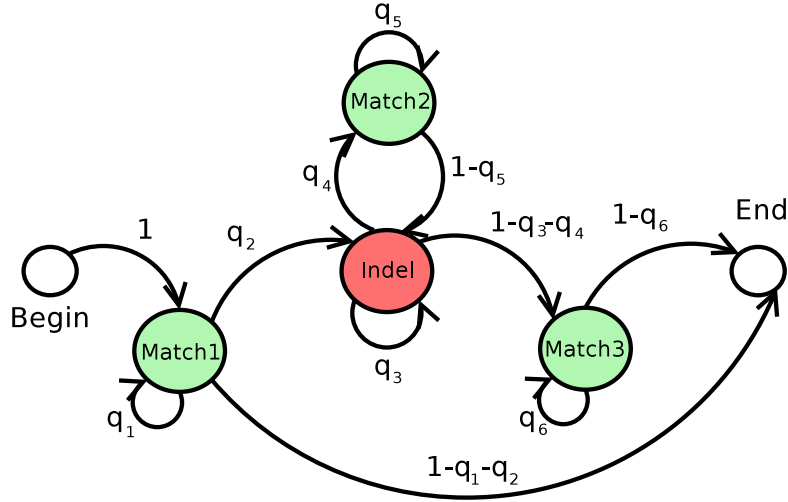}
    \label{fig.markovchain}}
   \vspace{-0.2in}
   \caption{Standard pair HMM corresponding to local Smith-Waterman
     alignments and the Markov Chain whose generative statistics,
     after Viterbi training, approximate the Viterbi statistics of the
     pair HMM for local alignments.}
\end{center}          

\vspace{-0.2in}
\end{figure*}

\begin{figure}[!tpb]
\label{fig.globalmarkov}
\centerline{
            \includegraphics[width=0.55\columnwidth]{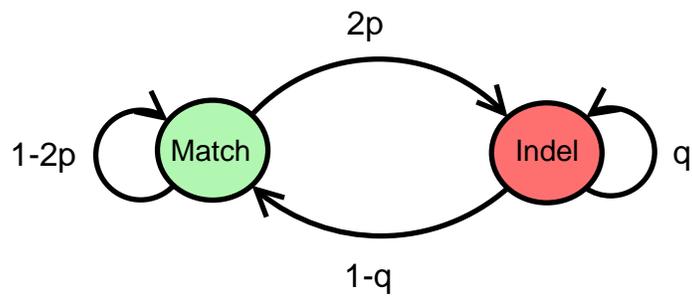}}
\caption{Markov chain for global alignments. See
  \cite{Schoenhuth10} for details.}
\end{figure}

In the following we only treat the more complex case of local
Smith-Waterman alignments. See \cite{Schoenhuth10} for the case of
global Needleman-Wunsch like alignments and Fig.~\ref{fig.globalmarkov}
for a picture of the corresponding Markov chain. For a treatment
of Needleman-Wunsch aligments the
Markov chain in Fig.~\ref{fig.globalmarkov} has to be, mutatis
mutandis, plugged into the computations of the subsequent
subsections.\par A local Smith-Waterman alignment with affine gap
penalties of two sequences $x=x_1...x_w,y=y_1...y_z$ is associated
with the most likely sequence of hidden states (i.e.~the {\it
  Viterbi path}) in the pair HMM of Fig.~\ref{fig.pairhmm}
\cite{Durbin}. The path of hidden states translates to an alignment of
the two sequences by emitting the necessary symbols along the
run. Statistics on Viterbi paths in HMMs pose hard mathematical
problems and have not been fully understood. In analogy to
\cite{Schoenhuth10}, we construct a Markov chain whose common,
generative statistics mimick the Viterbi statistics of interest
here. Hence probabilities derived from this Markov chain serve as
approximations of (\ref{eq.milp}). We do this by the following steps:
\begin{enumerate}
\item We take the Markov chain of the pair HMM in
  Fig.~\ref{fig.pairhmm} as a template.
\item We add two match states $M1$, $M3$. The original match state
  is $M2$.
\item We merge the initial resp.~terminal regions into one start
  resp.~end state.
\item We collapse states $X$ and $Y$ into one indel state $I$.
\end{enumerate}
\begin{sloppypar}

The Markov chain approach is justified by the fact that consecutive
runs in Viterbi paths are approximately governed by the {\em geometric
  distribution} which is precisely what a Markov chain reflects. To
see this note that to stay with a state in a Viterbi path is,
approximately, associated with that a self-transition attains maximum
probability in the next step. This depends both on the original
transition probability and the backward probability which depends on
the observed subsequence to follow starting from that state (see
\cite{Ephraim02}, (4.30) and the related discussion). Since it is a
general, computational assumption that the background distribution on
observed symbols (amino acids) is position-independent Viterbi path
transitions can be assumed to be (approximately) position-independent,
too. Note that assuming sequence to be position-independent is
reflected by that scoring schemes are position-independent. Clearly,
this is a computational assumption---we are aware of that the
biological reality can be different.\par The second point is to take
into account the non-stationary character of the original Markov
chain. Note that in local alignments, initial and final consecutive
stretches of (mis)matches are longer than intermediate (mis)match
stretches which translates to $q_1,q_6>q_5$ in Fig.~\ref{fig.pairhmm}.
To see this in more detail, note first that the related discussion in
\cite{Ephraim02} is on {\em stationary} HMMs. The non-stationarity of
the pair HMM under consideration here is due to that the initial and
terminal regions are heavily position-dependent. As a result, the
Viterbi paths under consideration have a {\em memory} which can
contradict the Markov assumption. The most striking effect is that
\begin{equation}
\Prob(X_{t+1}=I\,|\,X_t=M,X_{t-1}\in\{M,I\}) > 
\Prob(X_{t+1}=I\,|\,X_t=M,X_{t-1}=RY1).
\end{equation}
which reflects that to open up a gap shortly after having initiated
the core alignment tends to be avoided in order to circumvent an early
gap penalty. In symbols, this means that it is more likely to postpone
the core alignment and see (RY1)(RY1)(RY1) (and, possibly, some more
(RX1) before that) than running into an early gap after alignment
initiation (RY1)MI. Similar considerations hold for the terminal
regions. The second point addresses this by adding initial and
terminal match regions M1 and M3 which take the non-stationary
character of these areas into account. Point $3$ merely reflects that
we are only interested in statistics on alignment regions. Point $4$
finally accounts for that we do not make a difference between
insertions and deletions due to the involved symmetry (relative to
exchanging sequences).
\end{sloppypar}

\subsection{Algorithmic Solution of the MILP}
\label{ssec.algo}

We define $C_{n,k,d}$ to be the set of sequences over the alphabet
$\mathrm{B,M1,I,M2,M3,E}$ (for Begin, Match1, Indel, Match2, Match3
and End) of length $n$ that contain at least $d$ consecutive $I$
stretches of length at least $k$. Let $A_n:=\{X_n={\rm
  M3},X_{n+1}={\rm E}\}$ be the set of sequences with an alignment
region of length $n$. We then suggest the following procedure to
compute approximations of the probabilities (\ref{eq.milp}) where
$T(\sigma_1,\sigma_2)$ is supposed to be a pool of protein pairs $(x,y)$ whose alignments exhibit
alignment similarity $\Sim(x,y)\in [\sigma_1,\sigma_2]$.\\

{\bf 1:} Compute alignments for all sequence pairs in $T(\sigma_1,\sigma_2)$.\par
{\bf 2:} Infer parameters $q_1,q_2,q_3,q_4,q_5,q_6$ of the Markov chain by 
       Viterbi training it with the alignments.\par
{\bf 3:} $n\leftarrow$ length of the alignment of $x$ and $y$\par
{\bf 4:} Compute $\Prob(C_{n,k,d}\cap A_n)$ as well as $\Prob(A_n)$, the probabilities that 
       the Markov chain 
       of Fig.~\ref{fig.markovchain} generates sequences from 
       $C_{n,k,d}\cap A_n$ and $A_n$\par
{\bf 5:} {\em Output} 
       \begin{equation}
       \Prob(C_{n,k,d}\,|\,A_n)=\frac{\Prob(C_{n,k,d}\cap A_n)}{\Prob(A_n)}
       \end{equation}
       as an {\em approximation} for (\ref{eq.milp}).\par

\begin{sloppypar}
The idea of step $1$ and $2$ is to specifically train the Markov chain
to generate alignments from the pool $T(\sigma_2,\sigma_2)$. In our
setting, Viterbi training translates to counting $M_1$-to-$M_1$,
$M_1$-to-$I$, $I$-to-$I$, $I$-to-$M_2$, $M_2$-to-$M_2$ and
$M_3$-to-$M_3$ transitions in the alignments under consideration to
provide maximum likelihood estimates for $q_1,q_2,q_3,q_4,q_5$ and
$q_6$.\end{sloppypar}

\subsection{Efficient Computation of $\Prob(C_{n,k,d}\cap A_n)$}
\label{ssec.effcomp}

The problem of computing probabilities of the type (\ref{eq.milp}) 
has been made the problem of computing the probability that the Markov
chain generates sequences from $C_{n,k,d}\cap A_n$ and $A_n$. While computing
\begin{equation}
\Prob(A_n) = \Prob(X_n = {\rm M_3})\cdot \Prob(X_{n+1}= {\rm E}\,|\,X_n={\rm M_3})
\end{equation}
is an elementary computation, the question of efficient computation
and/or closed formulas for probabilities of the type
$\Prob(C_{n,k,d}\cap A_n)$ had not been addressed in the mathematical
literature and poses a last, involved problem.\par The approach taken
here is related to the one taken in \cite{Schoenhuth10}, which treated
the special case of single consecutive runs (i.e.~$d=1$) in the
context of the two-state Markov chains which reflect null models for
global alignments. We generalize this in two aspects. First, we provide a
solution for more than two states (our approach applies for arbitrary
numbers of states). Second, we show how to deal with multiple
runs.\par The probability event design trick inherent to our solution
was adopted from that of \cite{Pekoz95}. 
The solution provided in \cite{Pekoz95} can be used for the (rather
irrelevant) case of global alignments with linear gap penalties,
i.e.~gap opening and extension are identically scored.  See also
\cite{Fu94,Bassino08} for related mathematical treatments of the
i.i.d. case.

In the following, let $i,j\in\{\mathrm{B,M_1,I,M_2,M_3,E}\}$ be indices
ranging over the alphabet of Markov chain states. Let $e_i\in\R^6$ be
the standard basis vector of $\R^6$ having a $1$ in the i-th component
and zero elsewhere. For example,
$e_{\rm I}=(0,0,1,0,0,0),e_{{\rm M_3}}=(0,0,0,0,1,0)$.  We furthermore denote the
standard scalar product on $\R^6$ by $\langle .\,,. \rangle$.\par
Efficient computation of the probabilities $\Prob(C_{n,k,d}\cap A_n)$
is obtained by a dynamic programming approach. As usual, we collect
the Markov chain parameters (in accordance with
Fig.~\ref{fig.markovchain}) into a state transition probability matrix
\begin{multline}
P = (p_{ij}:=\Prob(X_t=i\,|\,X_{t-1}=j))_{i,j\in\{\mathrm{B,M1,I,M2,M3,E}\}} = \\
\left( \begin{array}{cccccc}
0&0&0&0&0&0 \\
1&q_1&0&0&0&0 \\
0&q_2&q_3&1-q_5&0&0\\
0&0&q_4&q_5&0&0\\
0&0&0&0&q_6&0\\
0&1-q_1-q_2&1-q_3-q_4&0&1-q_6&1
\end{array}\right)
\end{multline}
and an initial probability distribution vector
$\pi=e_B=(1,0,0,0,0,0)^T$.  The initial distribution reflects that we
start an alignment from the 'Begin' state. More formally,
$\Prob(X_0={\rm B})=1$. For example, according to the laws that govern
a Markov chain, the probability of being in the indel state I at
position $t$ in a sequence generated by the Markov chain is
\begin{equation}
\Prob(X_t = {\rm I}) = \langle e_I,P^t\pi\rangle = \langle e_I,P^te_B\rangle.
\end{equation}
It can be seen that naive approaches to computing $\Prob(C_{n,k,d}\cap
A_n)$ result in runtimes that are exponential in $n$, the length of
the alignments, which is infeasible.  Efficient computation of
these probabilities is helped by adopting the event design trick
of \cite{Pekoz95}. In detail, we define
\begin{equation}
D_{t,k} := \{X_t = {\rm I},...,X_{t+k-1}={\rm I},X_{t+k}\ne {\rm I}\}
\end{equation}
to be the set of sequences that have a run of state I of length $k$
that stretches from positions $t$ to $t+k-1$ and ends at position
$t+k-1$, that is, the run is followed by a visit of state different
from I at position $t+k$.\par 
We further define 
\begin{equation}
\pi_{\rm I} := \frac{1}{(1-p_{{\rm I}\,{\rm I}})}\cdot (p_{{\rm
    BI}},p_{{\rm M_1I}},0,p_{{\rm M_2I}},p_{{\rm M_3I}},p_{{\rm
    EI}})^T
\end{equation}
which can be interpreted as the state the Markov chain is in if we
know that the Markov chain has left state I at the time step
before. Consider $\Prob(X_{t+s}={\rm I}\,|\,X_{t-1}={\rm I},X_t\ne
{\rm I})$ as the probability that the Markov chain is in state I at
period $t+s$ after having been in the state $\pi_{\rm I}$ at period
$t$ (note that this probability is independent of $t$ as we deal with
a homogeneous Markov chain).  Similarly $\Prob(A_{t+k+s}\,|\,D_{t,k})$
is the probability that the Markov chain transits from state ${\rm
  M_3}$ to state E at position $t+k+s+1$ while it has a run of state I
of length $k$ that stretches from positions $t$ to $t+k-1$ and ends at
position $t+k-1$.  Lastly, we introduce the variables
\begin{eqnarray}
Q_{l,m} & := &  \sum_{1\le s_1,...,s_m\le l\atop s_1+...+s_m = l}
\Prob(X_{s_1}={\rm I})\prod_{i=2}^{m}\Prob(X_{t+s_i}={\rm I}\,|\,X_{t-1}={\rm I},X_t\ne {\rm I})\\
R_{L,m} & := & \sum_{l=m}^{L}Q_{l,m}\Prob(A_{t+k+L-l}\,|\,D_{t,k}),\quad 1\le m\le L\le n
\end{eqnarray}
for $1\le m\le l\le n$ where the sum reflects summing over partitions
of the integer $l$ into $m$ positive, not necessarily different,
integers $s_i$. We then obtain the following lemma a proof of which
needs a generalized inclusion-exlusion argument. 

\begin{lemma}
\label{l.lemma1}
\begin{equation}
\Prob(C_{n,k,d}\cap A_n) 
= \sum_{m=1}^{\lfloor\frac{n}{k+1}\rfloor}(-1)^{m+d}\binom{m-1}{d-1}\cdot {(p_{{\rm II}}^{k-1}(1-p_{\rm II}))}^m\cdot R_{n-mk,m}.
\end{equation} 
\end{lemma}

{\bf Proof}: We start by transforming
\begin{equation}
r_{\rm B}(s) := \Prob(X_s={\rm I}) = \Prob(X_s={\rm I}\,|\,X_0 = {\rm B}) =
\langle e_{\rm I},P^s\pi\rangle, \quad s\ge 1
\end{equation}
and
\begin{equation}
\begin{split}
\label{eq.r3}
r&_{\rm I}(s) := \Prob(X_{t+s}={\rm I}\,|\,X_{t-1}={\rm I},X_t\ne {\rm I}) = 
\frac{\Prob(X_{t-1}={\rm I},X_t\ne {\rm I},X_{t+s}={\rm I})}{\Prob(X_{t-1}={\rm I},X_t\ne {\rm I})}\\
&= \frac{\sum_{i\ne {\rm I}}\Prob(X_{t-1}={\rm I},X_t=i,X_{t+s}={\rm I})}{\sum_{i\ne {\rm I}}\Prob(X_{t-1}={\rm I},X_t=i)}
= \frac{\Prob(X_{t-1}={\rm I})\sum_{i\ne {\rm I}}\Prob(X_t=i,X_{t+s}={\rm I}\,|\,X_{t-1}={\rm I})}
{\Prob(X_{t-1}={\rm I})\sum_{i\ne {\rm I}}\Prob(X_t=i\,|\,X_{t-1}={\rm I})}\\
&=\frac{\Prob(X_{t-1}={\rm I})\sum_{i\ne {\rm I}}\Prob(X_t=i\,|\,X_{t-1}={\rm I})\Prob(X_{t+s}={\rm I}\,|\,X_t=i)}
{\Prob(X_{t-1}={\rm I})\sum_{i\ne {\rm I}}\Prob(X_t=i\,|\,X_{t-1}={\rm I})}\\
&=\frac{\sum_{i\ne {\rm I}}p_{i{\rm I}}\langle e_{\rm I},P^se_i\rangle}
{\sum_{i\ne {\rm I}}p_{i{\rm I}}}
= \frac{1}{1-p_{{\rm I}{\rm I}}}\langle e_{\rm I},P^s\sum_{i\ne {\rm I}}p_{i{\rm I}}e_i\rangle
=\langle e_{\rm I},P^s\pi_{\rm I}\rangle
\end{split}
\end{equation}
According to elementary Markov chain theory, one obtains, where here
and in the following $a(k):=p_{{\rm I}{\rm I}}^{k-1}(1-p_{{\rm I}{\rm
    I}})$
\begin{equation}
\begin{split}
\label{eq.D1}
\Prob(D_{t,k}) &= \Prob(X_t={\rm I},...,X_{t+k-1}={\rm I},X_{t+k}\ne {\rm I})\\ 
&= \Prob(X_t={\rm I})
\cdot \prod_{i=1}^{k-1}\Prob(X_{t+i}={\rm I}\,|\,X_{t+i-1}={\rm I})
\cdot \Prob(X_{t+k}\ne {\rm I}\,|\,X_{t+k-1}={\rm I})\\
&= r_{\rm B}(t)\cdot p_{{\rm I}{\rm I}}^{k-1}\cdot (1-p_{{\rm I}{\rm I}})
\end{split}
\end{equation}
and similarly, for $t_2\ge t_1$
\begin{equation}
\begin{split}
\label{eq.D2}
\Prob(D_{t_2,k}\,|\,D_{t_1,k}) &= \Prob(D_{t_2,k}\,|\,X_{t_1+k-1} = {\rm I},X_{t_1+k}\ne {\rm I})\\ 
&= 
\begin{cases}
r_{\rm I}(t_2-t_1-k)p_{{\rm I}{\rm I}}^{k-1}(1-p_{{\rm I}{\rm I}}) & t_2-k>t_1\\
0 & t_2-k\le t_1
\end{cases}
\end{split}
\end{equation}
Plugging (\ref{eq.D1}) and (\ref{eq.D2}) together yields, for $1\le t_1<...<t_m\le n-k+1$,
\begin{multline}
\begin{split}
\label{eq.D3}
\Prob(D_{t_1,k}\cap ... \cap D_{t_m,k}) 
= 
\begin{cases}
  \Prob(D_{t_1,k})\cdot\prod_{i=1}^{m-1}\Prob(D_{t_{i+1}}\,|\,D_{t_i}) & \forall i: t_{i+1}-t_i > k\\
  0 & \text{else}
\end{cases}\\
=                                  
\begin{cases}
  r_{\rm B}(t_1)\cdot\prod_{i=1}^{m-1}r_{\rm I}(t_{i+1}-t_i-k)\cdot (p_{{\rm I}{\rm I}}^{k-1}(1-p_{{\rm I}{\rm I}}))^m & \forall i: t_{i+1}-t_i > k\\
  0 & \text{else}
\end{cases}\\
\end{split}
\end{multline}
Including this into the definition of the $Q_{l,m}$ and $R_{L,m}$ yields
\begin{equation}
Q_{l,m} = \sum_{1\le s_1,...,s_m\le
  l\atop s_1+...+s_m = l}r_{\rm B}(s_1)\prod_{i=2}^{m}r_{\rm I}(s_i),\quad 1\le m\le l\le n
\end{equation}
and 
\begin{equation}
R_{L,m} = \sum_{l=m}^LQ_{l,m}r_{\rm M_3}(s_{L-l}).
\end{equation}
We now observe that
\begin{equation}
\label{eq.Cdnk}
C_{n,k,d} = \cup_{1\le t_1<t_2<...<t_d\le n-k+1}(D_{t_1,k}\cap ... \cap D_{t_d,k}).
\end{equation}
and we recall that we would like to compute 
\begin{equation}
\Prob(C_{n,k,d}\cap A_n)
\end{equation}
where $A_n := \{X_n = {\rm M_3}, X_{n+1} = {\rm E}\}$
is the set of sequences that have an alignment region of length
$n$. Proceeding by inclusion-exclusion yields
\begin{equation}
\begin{split}
\label{eq.CdnkAn}
\Prob(C_{n,k,d}\cap A_n) &\stackrel{(\ref{eq.Cdnk})}{=} \Prob(\cup_{1\le t_1<t_2<...<t_d\le n-k+1}(D_{t_1,k}\cap ... \cap D_{t_d,k}\cap A_n))\\
&= \sum_{m=d}^{\lfloor\frac{n}{k+1}\rfloor}K_{m,d}\cdot\Prob(D_{t_1,k}\cap ... \cap D_{t_m,k}\cap A_n)\\
&= \sum_{m=d}^{\lfloor\frac{n}{k+1}\rfloor}K_{m,d}\cdot\Prob(D_{t_1,k}\cap ... \cap D_{t_m,k})\cdot \Prob(A_n\,|\,D_{t_m,k})
\end{split}
\end{equation} 
where $\lfloor\frac{n}{k+1}\rfloor$ reflects the number of non-overlapping
events $D_{t_i}$, representing subsequences of length $k+1$, that fit
into a sequence of length $n$ and
\begin{equation}
\label{eq.kmd}
K_{m,d}=(-1)^{m+d}\binom{m-1}{d-1}
\end{equation}
is a generalized inclusion-exclusion coefficient. While the result can
be obtained from considerations that are analogous to that of the
usual case $d=1$ (note that $K_{m,1}=(-1)^{m+1}$ just results in the
usual inclusion-exclusion), it is not common in the mathematical
literature. See the subsequent lemma \ref{l.kmd} for a formal statement
and a proof.\par We further define, by computations that
are similar to (\ref{eq.r3}),
\begin{multline}
r_{\rm M_3}(s) := \Prob(A_{t+k+s}\,|\,D_{t,k}) \\
= \Prob(X_{t+k+s} = {\rm M_3}\,|\,D_{t,k})\cdot \Prob(X_{t+k+s+1}={\rm E}\,|\,X_{t+k+s}={\rm M_3}) 
= \langle e_{\rm M_3},P^s\pi_{\rm I}\rangle\cdot p_{{\rm M_3}{\rm E}}.
\end{multline}
By computations that are analogous to those of \cite{Schoenhuth10},
where in the following $a(k):=p_{{\rm I}{\rm I}}^{k-1}(1-p_{{\rm
    I}{\rm I}})$
\begin{equation}
\label{eq.final}
\begin{split}
\Prob&(C_{n,k,d}\cap A_n) \stackrel{(\ref{eq.CdnkAn})}{=} 
\sum_{i=d}^mK_{m,d}\Prob(D_{t_1,k}\cap ... \cap D_{t_m,k})\cdot \Prob(A_n\,|\,D_{t_m,k})\\
&\stackrel{(\ref{eq.D3})}{=} \sum_{m=1}^{\lfloor\frac{n}{k+1}\rfloor}K_{m,d}\cdot a(k)^m
\sum_{1\le t_1<...<t_m\le n-k+1\atop t_{i+1}-t_i>k}
r_1(t_1)\cdot\prod_{i=1}^{m-1}r_{\rm I}(t_{i+1}-t_i-k)\cdot r_{\rm M_3}(n-t_m-k)\\
&= \sum_{m=1}^{\lfloor\frac{n}{k+1}\rfloor}K_{m,d}\cdot a(k)^m
\left[\sum_{l=m}^{n-mk}Q_{l,m}\cdot r_{\rm M_3}(n-mk-l))\right].\\
&= \sum_{m=1}^{\lfloor\frac{n}{k+1}\rfloor}K_{m,d}\cdot a(k)^m\cdot R_{n-mk,m}\\
&= \sum_{m=1}^{\lfloor\frac{n}{k+1}\rfloor}(-1)^{m+d}\binom{m-1}{d-1}\cdot {(p_{{\rm I}{\rm I}}^{k-1}(1-p_{{\rm I}{\rm I}
}))}^m\cdot R_{n-mk,m}.
\end{split}
\end{equation} 
\qed

\begin{lemma}
\label{l.kmd}
Let $D_i,i\in \{1,...,N\}$ be a family of $N$ events. Then it holds
that
\begin{equation}
\Prob(\cup_{1\le i_1<...<i_d\le N}(D_{t_1}\cap ... \cap D_{t_d}) 
= \sum_{m=d}^N(-1)^{m+d}\binom{m-1}{d-1}\sum_{1\le i_1<...<i_m\le N}\Prob(D_{i_1}\cap ... \cap D_{i_m}).
\end{equation}
\end{lemma}


{\bf Proof.} The proof proceeds similarly to that of the special, well
known case of $d=1$. 
Let $\omega\in\cup_{i_1<...<i_d}(D_{i_1}\cap ... \cap D_{i_d})$ such
that, w.l.o.g., $\omega$ is contained in $D_1,...,D_n$ where $n\le N$,
but not contained in $D_{n+1},...,D_N$. Let $\mathbf{1}_{\omega}$ be
the indicator function of $\omega$. According to the choice of
$\omega$ it holds that
\begin{equation}
\mathbf{1}_{\omega}(D_i) = \begin{cases} 1 & 1\le i\le n\\
                                     0 & \text{else}
                       \end{cases}.
\end{equation}
Proceeding along the lines of the proof of the usual inclusion-exclusion theorem
($d=1$) (see e.g. \cite{Brualdi}) it suffices to show that
\begin{equation}
\mathbf{1}_{\omega}(\cup_{i_1<...<i_d}(D_{t_1}\cap ... \cap D_{t_d})) = 
\sum_{m=d}^n(-1)^{m+d}\binom{m-1}{d-1}\sum_{J\subset\{1,...,n\}\atop |J|=l}
\mathbf{1}_{\omega}(\cap_{j\in J} D_j)
\end{equation}
Evaluating this equation at $\omega$ amounts to showing that
\begin{equation}
1 = \sum_{m=d}^n(-1)^{n+d}\binom{m-1}{d-1}\binom{n}{m}.
\end{equation}
This is done by induction on $d$. The case $d=1$ 
\begin{equation}
1 = \sum_{l=1}^n(-1)^{n+1}\binom{m-1}{0}\binom{n}{m} = \sum_{m=1}^n(-1)^{n+1}\binom{n}{m}
\end{equation}
is the usual case of standard inclusion-exclusion which, by putting
the right hand side to the left, follows from
\begin{equation}
\label{eq.d=1}
\sum_{m=0}^n(-1)^n\binom{n}{m} = (1-1)^n = 0.
\end{equation}
$d\to d+1$: In this case, 
In this case, we have to show that
\begin{equation}
\sum_{m=d+1}^n(-1)^{n+d+1}\binom{m-1}{d}\binom{n}{m} = 1
\end{equation}
Therefore, we can assume that
\begin{equation}
\label{eq.iv}
\sum_{m=d}^n(-1)^{m+d}\binom{m-1}{d-1}\binom{n}{m} = 1\quad\Leftrightarrow\quad
\sum_{m=d+1}^n(-1)^{l+d+1}\binom{m-1}{d-1}\binom{n}{m} = \binom{n}{d} - 1.
\end{equation}
Furthermore, it holds that
\begin{equation}
\label{eq.lknl}
\binom{m}{d}\binom{n}{m} = \frac{n!}{m!(n-m)!}\cdot\frac{m!}{d!(m-d)!}
= \frac{n!}{d!(n-d)!}\cdot\frac{(n-d)!}{(m-d)!(n-m)!} = \binom{n}{d}\binom{(n-d)}{(m-d)}. 
\end{equation}
We proceed
\begin{equation}
\begin{split}
\sum_{m=d+1}^n&(-1)^{n+d+1}\binom{m-1}{d}\binom{n}{m} = \sum_{m=d+1}^n(-1)^{n+d+1}[\binom{m}{d}-\binom{m-1}{d-1}]\binom{n}{m}\\
&\stackrel{(\ref{eq.iv})}{=} [\sum_{m=d+1}^n(-1)^{n+d+1}\binom{m}{d}\binom{n}{m}] - \binom{n}{d} + 1\\
&\stackrel{(\ref{eq.lknl})}{=} [\sum_{m=d+1}^n(-1)^{n+d+1}\binom{n}{d}\binom{(n-d)}{(m-d)}] - \binom{n}{d} + 1\\
&= \binom{n}{d}[\sum_{m=d+1}^n(-1)^{n+d+1}\binom{(n-d)}{(m-d)}] - \binom{n}{d} + 1\\
&= \binom{n}{d}[\sum_{m'=1}^n(-1)^{n+1}\binom{n}{m'}] - \binom{n}{d} + 1\\
&\stackrel{(\ref{eq.d=1})}{=} \binom{n}{d} - \binom{n}{d} + 1 = 1
\end{split}
\end{equation}
which concludes the proof.\qed\\

The consequences can be summarized in the
following theorem.

\begin{theorem}
A full table of values $\Prob(C_{n,k,d}\cap A_n),k\le n\le N$
can be computed in $O(N^3)$ runtime.
\end{theorem}

{\bf Proof}. 
Observing the recursive relationship
\begin{equation}
\label{eq.qlm}
  Q_{l,m} = \sum_{s=1}^{l-m+1}\Prob(X_{t+s}={\rm I}\,|\,X_{t-1}={\rm I},X_t\ne {\rm I})Q_{l-s,m-1},\quad m>1
\end{equation}
yields a standard dynamic programming procedure by which the ensemble
of the $Q_{l,m}$ and the $R_{L,m}$ ($1\le m\le l,L\le N$) can be
computed in $O(N^3)$ runtime. This also requires that the values
$\Prob(X_{s}={\rm I}),\Prob(X_{t+s}={\rm I}\,|\,X_{t-1}={\rm I},X_t\ne {\rm I})$ have been
precomputed which can be done in time linear in $N$. After
computation of the $Q_{l,m}$ and the $R_{L,m}$, computation of the
$\Prob(C_{n,k,d}\cap A_n),1\le k\le n\le N$ then equally requires
$O(N^3)$ time which follows from lemma \ref{l.lemma1}.\qed

\section{Results}
\label{sec.results}
\begin{table*}[!t]
{\footnotesize
\caption{Markov chain parameters for local alignments.\label{tab.markovlocal}}
\begin{center}
\begin{tabular}{c|cccccccc}
\hline
\multicolumn{9}{c}{{\bf Twilight Zone (Twi)}}\\
\hline
Similarity (\%)&20 - 30 & 30 - 40 & 40 - 50 & 50 - 60 & 60 - 70 & 70 - 80 & 80 - 90 & 90 - 100\\ 	
\hline 
No.~Alignments & - & 27	& 1896	& 12512	& 9716	& 3956	& 1733	& 259\\
$q_1$	& - & 0.9552 & 0.9606 & 0.9564 & 0.9485 & 0.9364 & 0.9188 & 0.9149\\
$q_2$	& - & 0.0216 & 0.0300 & 0.0315 & 0.0265 & 0.0167 & 0.0078 & 0.0036\\
$q_3$	& - & 0.7500 & 0.6667 & 0.5893 & 0.4692 & 0.3210 & 0.3640 & 0.0000\\
$q_4$	& - & 0.0588 & 0.1948 & 0.2185 & 0.1739 & 0.0979 & 0.0459 & 0.0000\\
$q_5$	& - & 0.8261 & 0.9439 & 0.9353 & 0.9226 & 0.8999 & 0.9253 & 1.0000\\
$q_6$	& - & 0.9417 & 0.9514 & 0.9472 & 0.9335 & 0.9077 & 0.8991 & 0.7755\\
\hline
\hline
\multicolumn{9}{c}{{\bf Superfamilies (Sup)}}\\
\hline
Similarity (\%)&20 - 30 & 30 - 40 & 40 - 50 & 50 - 60 & 60 - 70 & 70 - 80 & 80 - 90 & 90 - 100\\ 	
\hline 
No.~Alignments & - & 44 & 3743 & 23726 & 18633 & 7275 & 2613 & 454\\
$q_1$ & - & 0.9511 & 0.9584 & 0.9568 & 0.9534 & 0.9528 & 0.9346 & 0.9273\\
$q_2$ & - & 0.0267 & 0.0330 & 0.0330 & 0.0266 & 0.0160 & 0.0085 & 0.0034\\
$q_3$ & - & 0.7643 & 0.6829 & 0.6043 & 0.5001 & 0.4044 & 0.2553 & 0.0000\\
$q_4$ & - & 0.0828 & 0.1962 & 0.2390 & 0.2390 & 0.2220 & 0.0959 & 0.0000\\
$q_5$ & - & 0.8952 & 0.9430 & 0.9410 & 0.9466 & 0.9674 & 0.9820 & 0.0000\\
$q_6$ & - & 0.9438 & 0.9508 & 0.9495 & 0.9443 & 0.9504 & 0.9407 & 0.7921\\
\hline
\end{tabular}\end{center}}{}
\end{table*}

\begin{table}[!t]
{\footnotesize
\caption{Markov chain parameters for global alignments.\label{tab.markovglobal}}
\begin{center}
\begin{tabular}{c|cccccccc}
\hline
\multicolumn{9}{c}{{\bf Twilight Zone (Twi)}}\\
\hline
Similarity (\%)&20 - 30 & 30 - 40 & 40 - 50 & 50 - 60 & 60 - 70 & 70 - 80 & 80 - 90 & 90 - 100\\ 	
\hline 
No.~Alignments & 31 & 1811 & 9156 & 616 & 36 & 7 & 8 & - \\ 	  
$1-2p$ & 0.9092 & 0.9290 & 0.9287 & 0.9311 & 0.9528 & 0.9790 & 0.9939 & -\\
$q$ & 0.2615 & 0.1835 & 0.1475 & 0.0994 & 0.0619 & 0.0269 & 0.0364 & - \\
\hline
\hline
\multicolumn{9}{c}{{\bf Superfamilies (Sup)}}\\
\hline
Similarity (\%)&20 - 30 & 30 - 40 & 40 - 50 & 50 - 60 & 60 - 70 & 70 - 80 & 80 - 90 & 90 - 100\\ 	
\hline 
No.~Alignments & 44 & 2925 & 17127 & 3234 & 1304 & 454 & 39 & - \\ 	  
$1-2p$ & 0.9054 & 0.9277 & 0.9292 & 0.9421 & 0.9630 & 0.9788 & 0.9900 & - \\
$q$ & 0.2528 & 0.1876 & 0.1482 & 0.0980 & 0.0523 & 0.0257 & 0.0097 & - \\
\hline
\end{tabular}\end{center}}{}
\end{table}

\begin{figure*}[t]
\centering
\includegraphics[width=0.49\textwidth]{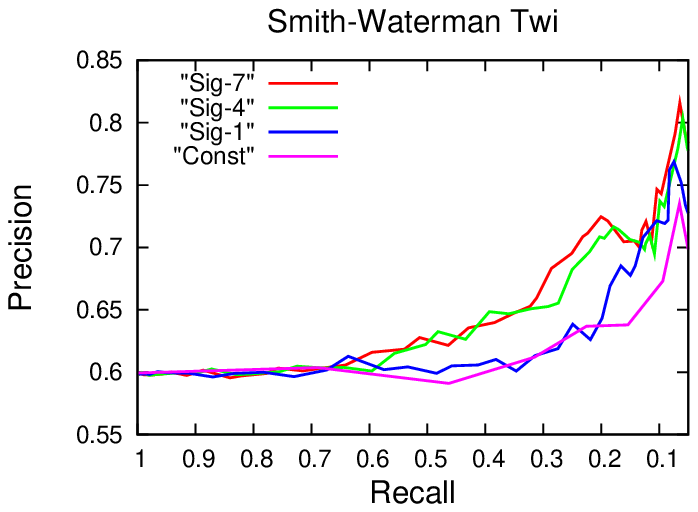}\;
\includegraphics[width=0.49\textwidth]{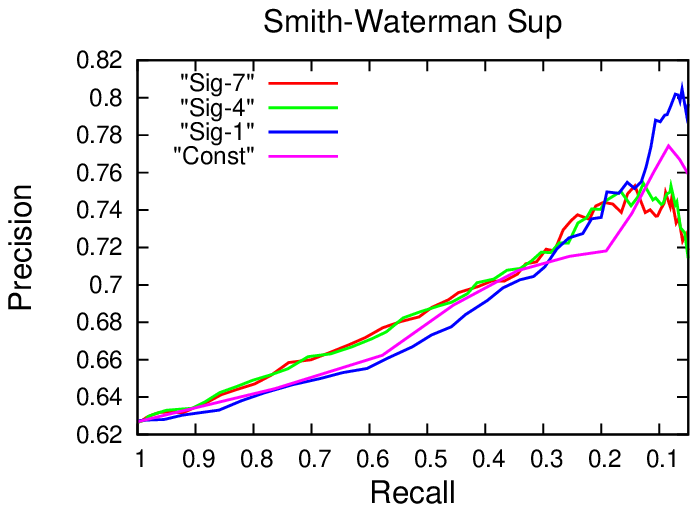}\\
\includegraphics[width=0.49\textwidth]{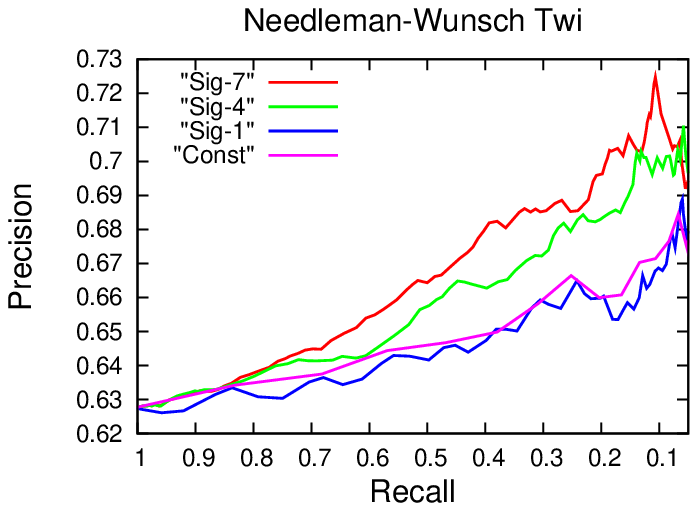}\;
\includegraphics[width=0.49\textwidth]{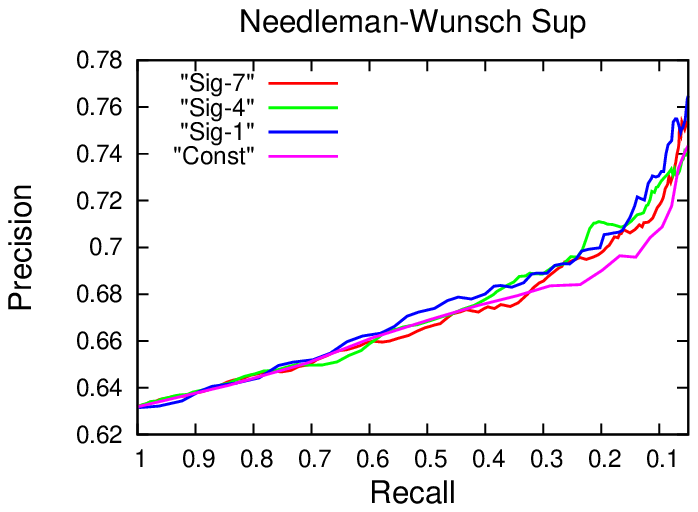}
\caption{Precision-Recall curves for the different sets of
  computational alignments. Recall is lowered through lowering the
  significance threshold $\theta$ for the strategies ${\rm
    Sig_D}(\theta)$ ($\theta=1.0$ for maximal recall of $1.0$) and
  for raising indel length in the baseline strategy Const (length =
  $1$ for maximal recall of $1.0$).}
\label{fig.plots}
\end{figure*}

\paragraph{Data}
\begin{sloppypar} 
We downloaded both the ``Superfamilies'' (Sup) and ``Twilight Zone''
(Twi) datasets together with their structural alignment information
from SABmark 1.65 \cite{VanWalle05}, including the suggested false
positive pairs (that is structurally unrelated, but apparently similar
sequences, see \cite{VanWalle05} for a detailed description).  While
Sup is a more benign set of structural alignments where protein pairs
can be assumed to be homologous and which contains alignments of up to
50\% identity, Twi is a worst case scenario of alignments between only
0-25 \% sequence identity where the presence of a common evolutionary
ancestor remains unclear.\par\end{sloppypar} To calculate pairwise
global resp.~local alignments we used the ``GGSEARCH''
resp.~''LALIGN'' tool from the FASTA sequence comparison package
\cite{Pearson88}. As a substitution matrix, BLOSUM50 (default) was
used.  GGSEARCH resp.~LALIGN implement the classical Needleman-Wunsch
(NW) resp.~Smith-Waterman (SW) alignment algorithm both with affine
gap penalties. We subsequently discarded global resp.~local alignments
of an e-value larger than $10.0$ resp.~$1.0$, as suggested as a default
threshold setting \cite{Pearson88}, in order to ensure to only treat
alignments which can be assumed not to be entirely random.\par We then
subdivided the resulting $4$ groups (NW Twi, NW Sup, SW Twi and SW
Sup) of computational alignments into pools of alignments of
similarity in $[\sigma,\sigma + 10]$ where $\sigma$ ranged from $20$
to $90$. We then trained parameters (using also the false positive
SABmark alignments in order to obtain unbiased null models) for the
$36 = 4\times 9$ different Markov chains ($2$-state as in
\cite{Schoenhuth10} resp.~$6$-state as described here for global
resp.~local) and computed probability tables as described in the
Methods section. After computation of probability tables, false
positive alignments were discarded. See below for Markov chain
parameters.\par The remaining (non
false-positive) NW Twi, NW Sup, SW Twi and SW Sup alignments contained
$179018$, $407629$, $20853$ and $86233$ gap positions contained in
$122701$, $276082$, $17776$ and $68513$ gaps. In the global alignments
this includes also initial and end gaps.

\subsection{Evaluation Strategies}
\label{ssec.strategies}

Based on efficient computation of probabilities of the type
(\ref{eq.milp}) we devise strategies 
${\rm Sig_D}(\theta)$
for predicting indel reliability in NW and SW alignments where
$D=1,4,7$. Let $K$ be the length of the $L$-th longest indel in the NW
resp.~SW alignment of proteins $x,y$. In strategy ${\rm
  Sig_D}(\theta)$, this indel is classified as reliable if
\begin{equation}
{\rm Sig_D}(\theta):\quad\Prob(I_{\min(D,L)}(x,y)\ge K\,|\,L(x,y),\Sim(x,y))\le \theta.
\end{equation}
In other words, we look up whether it is significant that an
alignment of length $L(x,y)$ and similarity $\Sim(x,y)$ contains at
least $L$ resp.~$D$, in case of $D>L$ resp.~$D\le L$, indels of size
$K$. Note that, since $L\ge 1$ hence $\min(D,L)=1$, in strategy ${\rm Sig_1}(\theta)$ an
indel of length $K$ is evaluated as reliable if and only if
the indel is significantly long without considering its
relationship with the other gaps in the alignment. This is different
for strategy ${\rm Sig_7}(\theta)$ where, for example, the $6$-th
longest indel is evaluated as reliable if it is significant to have at
least $6$ indels of that length ($\min(D,L)=6$) whereas the $8$-th
longest indel is supposed to be reliable if there are at least $7$
indels of that length ($\min(D,L)=7$). Note that in strategy ${\rm
  Sig_7}(\theta)$ already shorter indels are classified as
reliable in case that there are many indels of that length in the
alignment which is not the case in strategy ${\rm
  Sig_1}(\theta)$. Clearly, raising $D$ beyond $7$ might make
sense. For sake of simplicity only, we restricted our attention to
$D=1,4,7$.\par As a simple baseline method we suggest ${\rm Const}$
which considers an indel as reliable if its length exceeds a constant
threshold. Both raising the constant length threshold in ${\rm Const}$
and lowering $\theta$ in ${\rm Sig_D}(\theta)$ lead to reduced amounts
of indels classified as reliable.

\paragraph{Evaluation Measures}

\begin{table*}[t]
{
\begin{center}
{\begin{tabular}{c|c|c|c|c|c|c|c|c|}
\label{tab.gapfrac} 
& \multicolumn{2}{|c|}{NW Twi} & \multicolumn{2}{|c|}{NW Sup} & \multicolumn{2}{|c|}{SW Twi} & \multicolumn{2}{|c|}{SW Sup}\\
& $\le 4$ & $>30$ & $\le 4$ & $>30$ & $\le 4$ & $>20$ & $\le 4$ & $>20$\\
\hline
FGP & 0.42 & 0.03 & 0.41 & 0.01 & 0.63 & 0.01 & 0.53 & 0.02\\
\hline
PPV & 0.58 & 0.64 & 0.53 & 0.92 & 0.53 & 0.69 & 0.45 & 0.87\\
\hline
\end{tabular}}
\caption{\label{tab.basic} Fractions of Gap Positions (FGP) contained
  in gaps of different length ranges and Fractions of True Gap
  Positions (PPV) contained in such gaps}.
\end{center}
}
\end{table*}

We found that for both global and local alignments further evaluation
of gaps of length at most $4$ and length greater than $30$ (global)
resp.~$20$ (local) did not make much sense (see table
\ref{tab.gapfrac} for statistics).  However, for gaps of length
ranging from $5$ to $20$ resp.~$30$ in local resp.~global alignments a
significance analysis made sense.\par We evaluated the indel positions
in gaps of length $5-20$ resp.~$5-30$ in local resp.~global alignments
by defining a true positive (TP) to be a computational gap position
which is classified as reliable (meaning that it is found to be
significant by ${\rm Sig_D}(\theta), D=1,4,7$ or long enough by Const)
and coincides with a true structural indel position in the reference
structural alignment as provided by SABmark. Correspondingly, a false
positive (FP) is a gap position classified as reliable which cannot be
found in the reference alignment. A true negative (TN) is a gap
position not classified as reliable and not a structural indel
position and a false negative (FN) is not classified as reliable but
refers to a true structural indel position. Recall, as usual, is
calculated as $TP/(TP+FN)$ whereas Precision (also called PPV=Positive
Predictive Value) is calculated as $TP/(TP+FP)$.

\subsection{Discussion of Results}

\begin{table*}[t]
{\footnotesize
\begin{center}
{\begin{tabular}{c|cccc|cccc|cccc|cccc}

Recall & \multicolumn{3}{|c}{$-\log\,\theta$} & IL & \multicolumn{3}{|c}{$-\log\,\theta$} & IL 
& \multicolumn{3}{|c}{$-\log\,\theta$} & IL & \multicolumn{3}{|c}{$-\log\,\theta$} & IL \\
\hline
1.0 & 0.0 & 0.0 & 0.0 & 5 & 0.0 & 0.0 & 0.0 & 5 & 0.0 & 0.0 & 0.0 & 5 & 0.0 & 0.0 & 0.0 & 5\\
0.75 & 2.0 & 2.0 & 1.0 & 5 & 2.0 & 2.0 & 1.0 & 6 & 19.0 & 18.5 & 6.5 & 6 & 21.0 & 19.5 & 7.0  & 6\\
0.5  & 2.5 & 2.5 & 1.5 & 6 & 3.0 & 3.0 & 1.5 & 7 & 28.0 & 24.0 & 10.5& 8 & 30.0 & 27.0 & 11.5 & 8\\
0.25 & 3.5 & 3.5 & 2.5 & 8 & 4.5 & 4.5 & 3.0 & 10& 38.5 & 33.0 & 16.0& 11& 41.5 & 36.0 & 18.0 & 11\\
\hline
& ${\rm Sig_7}$ & ${\rm Sig_4}$ & ${\rm Sig_1}$ & C & ${\rm Sig_7}$ & ${\rm Sig_4}$ & ${\rm Sig_1}$ & C
& ${\rm Sig_7}$ & ${\rm Sig_4}$ & ${\rm Sig_1}$ & C & ${\rm Sig_7}$ & ${\rm Sig_4}$ & ${\rm Sig_1}$ & C\\
& \multicolumn{4}{|c|}{{\bf SW Twi}} & \multicolumn{4}{|c|}{{\bf SW Sup}} 
& \multicolumn{4}{|c|}{\bf NW Twi} & \multicolumn{4}{|c}{\bf NW Sup}\\
\end{tabular}}
\vspace{2ex}
\caption{\label{tab.support} Relationship between Recall and
  $\theta$ (displayed as $-\log\,\theta$) for strategies ${\rm Sig_D}$
  and indel length (= IL) for strategy Const (= C).}
\end{center}
}
\end{table*}


Results are displayed in Figure \ref{fig.plots} where we have plotted
Precision vs.~Recall while lowering $\theta$ for the strategies ${\rm
  Sig_D}(\theta)$ and increasing indel length for the baseline method
Const. While Recall $=1.0$ relates to $\theta=1.0$ in the strategies
${\rm Sig_D}$ maximal recall relates to indel length $5$ in the
strategy Const. Table \ref{tab.support} displays further supporting
statistics on the relationship between choices of $\theta$ resp.~indel
length and Recall.\par A first look reveals that indel reliability
clearly increases for increasing indel length---longer indels are more
likely to contain true indel positions. However, further improvements
can be achieved by classifying indels as reliable according to
significance. For the Sup alignments improvements over the baseline
method are only slight. For both local and global alignments strategy
${\rm Sig_1}$ is an option in particular when it comes to achieving
utmost precision which can be raised up to $0.8$. For the Twi
alignments differences are obvious. More importantly, just considering
indel length without evaluating multiplicity does not serve to achieve
substantially increased Precision. Here, multiplicity is decisive
which in particular confirms the findings on twilight zone alignments
reported in \cite{Polyanovsky08}. In the Twi alignments Precision can
be raised up to about $0.7$. Note that \cite{Schlosshauer02} achieve
$0.7$ Precision on both match and gap positions for structural
alignments (not from SABmark) of $25-30\%$ identity while reporting
that their evaluation does not work for alignments of less than $25\%$
identity which renders it not applicable for the Twi alignments.  The
posterior decoding aligner FSA which outperformed all other multiple
aligners in terms of Precision on both (mis)match and gaps in the
entire SABmark dataset, comprising both Sup and Twi \cite{Bradley09}
report Precision of $0.52$ (all other aligners fall below $0.5$)
without further re-evaluation of their alignments. This lets us
conclude that our statistical re-evaluation makes an interesting
complementary contribution to alignment re-evaluation.

\section{Conclusion}
\label{sec.conclusion}

Most recent studies have again pointed out that computational
alignments of all kinds need further re-evaluation in order to avoid
detrimental effects in downstream analyses of comparative genomics
studies. While exact gap placement is at the core of aligning sequence
positive prediction rates are worst within or closely around inferred
indels.  Here we have systematically addressed that indel size and
multiplicity can serve as indicators of alignment artifacts.
We have developed a pair HMM based statistical evaluation pipeline
which can soundly distinguish between spurious and reliable indels in
alignments with affine gap penalties by measuring indel significance
in terms of indel size and multiplicity. As a result we are able to
reliably identify indels which are more likely to enclose true
structural indel positions as provided by SABmark, raising positive
prediction rates up to $0.7$ even for worst-case twilight zone
alignments of maximal $25\%$ sequence identity. Since previous
approaches predominantly addressed re-evaluation of match/mismatch
positions we think that we have made a valuable, complementary
contribution to the issue of alignment re-evaluation. Future work
of ours is concerned with re-evaluation of pair HMM based posterior
decoding aligners which have proven to be superior over score-based
aligners in a variety of aspects.

\subsection*{Acknowledgements}

AS is funded by donation from David DesJardins,
Google Inc. We would like to thank Lior Pachter for helpful
discussions on Viterbi path statistics.

\bibliographystyle{plain}

\end{document}